\documentclass[reprint,
 amsmath,amssymb,
 aps,
prb,
superscriptaddress
]{revtex4-2}

\usepackage{graphicx}

\usepackage{dcolumn}
\usepackage{bm}
\usepackage[mathlines]{lineno}

\usepackage[dvipsnames]{xcolor}

\usepackage[
    colorlinks=true,
    linkcolor=blue,
    citecolor=blue,
    urlcolor=blue
]{hyperref}

\begin{document}
\preprint{APS/123-QED}

\title{Photoelectrical readout and Ramsey interferometry of single shallowly implanted NV centers in diamond}

\author{Ilia Chuprina}
\email{ilia.chuprina@uni-ulm.de}
\affiliation{Institute for Quantum Optics, Ulm University, D-89081 Germany}

\author{Christoph Findler}
\affiliation{Institute for Quantum Optics, Ulm University, D-89081 Germany}
\affiliation{Diatope GmbH, Buchenweg 23, D-88444 Ummendorf, Germany}

\author{Johannes Lang}
\affiliation{Institute for Quantum Optics, Ulm University, D-89081 Germany}
\affiliation{Diatope GmbH, Buchenweg 23, D-88444 Ummendorf, Germany}

\author{Petr Siyushev}
\affiliation{Institute for Materials Research (IMO), Hasselt University, Wetenschapspark 1, B-3590 Diepenbeek, Belgium}
\affiliation{IMOMEC division, IMEC, Wetenschapspark 1, B-3590 Diepenbeek, Belgium}
\affiliation{Institute for Quantum Optics, Ulm University, D-89081 Germany}

\author{Fedor Jelezko}
\affiliation{Institute for Quantum Optics, Ulm University, D-89081 Germany}
\affiliation{Center for Integrated Quantum Science and Technology (IQST), 89081 Ulm, Germany}

\date{\today}

\begin{abstract}
Photoelectrical readout of the electronic spin state of the nitrogen-vacancy (NV) center in diamond is attracting significant interest due to the numerous advantages it possesses compared with conventional fluorescence readout. The higher charge carrier rate compared to the photon rate and the integration of the detection scheme on a chip can significantly advance quantum sensing and computing with color centers in diamond. Until now, photoelectric readout has been performed on ensembles of NV centers or single NV centers deep in ultrapure diamond substrates. However, many applications require the artificial creation and precise placement of shallow NV centers, and photoelectric detection of such centers has been challenging. Here we demonstrate photoelectrical readout and coherent control of the electronic spin of implanted shallow ($\sim$10 nm) NV centers buried by diamond overgrowth. The photoelectrically measured Ramsey $T_2^*$ agrees with conventional fluorescence readout and shows no measurable dependence on the readout photocurrent, for both shallow implanted and deep ingrown NV centers. We further find that overgrowth improves photoelectric readout by suppressing the background photocurrent. These results establish photoelectric readout as a viable route to chip-integrated, electrically detected nanoscale sensing and to spin registers based on engineered shallow NV centers.
\end{abstract}

\maketitle


\section{\label{sec:intro}Introduction}
The nitrogen-vacancy (NV) center in diamond is a well-established solid-state spin system with long coherence times and room-temperature operation~\cite{balasubramanian2009ultralong,DOHERTY20131}. Conventionally, the electronic spin state of the negatively charged NV center, NV$^-$, is read out optically through its spin-dependent red-shifted fluorescence. Under sufficiently intense optical excitation, NV$^-$ and NV$^0$ can interconvert through a charge-cycle process that generates free carriers~\cite{aslam2013photo,siyushev2013optically}. Photoelectric (PE) readout exploits this spin-dependent charge conversion by collecting the photoexcited electrons and holes with electrodes, so that the amplified photocurrent provides the spin-readout signal~\cite{bourgeois2015photoelectric}. This electrode-based detection scheme is attractive for chip-integrated quantum sensing and quantum-information devices~\cite{bourgeois2020photoelectric}.

Many applications, however, require NV centers with controlled position and depth a few nanometers below the diamond surface. Such shallow NV centers are essential for nanoscale sensing of external spins and surface species, including nuclear magnetic resonance from molecules placed on the diamond surface~\cite{staudacher2013nuclear,muller2014nuclear,LoretzAPL2014,shi2015single,balasubramanian2008nanoscale,maze2008nanoscale}, entanglement-assisted quantum sensing \cite{zhou2025entanglement, rovny2025multi}, and quantum-information protocols based on implanted NV pairs or registers~\cite{neumann2010quantum,dolde2014high,joas2025high}. These requirements make shallow, artificially created NV centers the relevant platform for scalable nanoscale quantum devices for quantum sensing and computing.

So far, PE readout of single NV centers has mainly been demonstrated using deeply buried, naturally occurring or in-grown NV centers, where surface-related charge instability and implantation-induced damage are minimized~\cite{siyushev2019photoelectrical}. Extending PE readout to such shallow NV centers is therefore not a straightforward transfer, but a necessary step toward electrically detected nanoscale sensing and engineered quantum devices.

The main difficulty arises from the defect environment created by shallow implantation and the proximity to the diamond surface. Low-energy nitrogen implantation enables shallow NV formation, but the conversion yield is typically below $5\%$~\cite{pezzagna2010creation}, leaving a residual bath of substitutional nitrogen, $\mathrm{N_s}$~(P1), together with vacancy-related complexes~\cite{favaro2017tailoring}. These defects can act as charge traps and additional photoionization sources, thereby reducing the PE readout contrast and signal-to-noise ratio. In particular, $\mathrm{N_s}$ is a deep donor with an ionization threshold of $\sim 2.3\,\mathrm{eV}$, corresponding to wavelengths $\sim 540\,\mathrm{nm}$~\cite{iakoubovskii2000optical}. Since NV centers are commonly addressed using green excitation near $532\,\mathrm{nm}$, $\mathrm{N_s}$ can be photoionized together with the NV center~\cite{rosa1999photoionization}, producing parasitic photocurrent that can obscure the spin-dependent PE signal~\cite{bourgeois2017enhanced,siyushev2019photoelectrical}.

These limitations can be addressed by burying implanted NV centers below the surface while keeping them sufficiently shallow for near-surface sensing applications. One route to achieve this is indirect diamond overgrowth, where deterministically implanted NV centers are capped by an ultrapure diamond layer~\cite{findler2020indirect}. The final NV depth can be adjusted through the implantation energy and overgrowth thickness, allowing the trade-off between surface proximity and spin coherence to be optimized \cite{staudacher_APL2012}. In addition, overgrowth can modify the near-surface defect environment and improve the spin-bath composition around implanted NV centers. However, PE readout of shallow implanted NV centers after diamond overgrowth has not yet been demonstrated and investigated.

In this work, we demonstrate PE readout and coherent control of the electron spin of shallow implanted NV centers after diamond overgrowth. We compare the PE-Rabi contrast of single implanted NV centers with that of deeply buried naturally occurring NV centers and analyze the defect bath and background photoionization relevant for PE detection. Furthermore, we investigate the photoelectrically detected Ramsey dephasing time, $T_2^*$, as a function of laser intensity and photocurrent. The extracted $T_2^*$ values agree well with those obtained using conventional fluorescence readout, indicating that the NV ground-state spin dephasing is preserved in the presence of photocurrent-induced magnetic fields. These results establish PE readout as a viable approach for quantum sensing and quantum-information protocols with engineered shallow single NV centers.

\begin{figure}[h]
    \includegraphics[width=0.97\columnwidth]{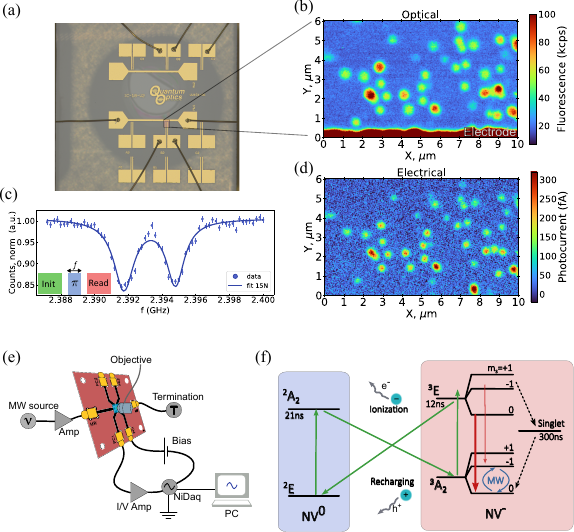}
    \caption{Main characteristics of photoelectric readout of implanted NV centers. (a) Microscope image of the diamond chip used for the experiments. The position of the electrodes coincides with the ion implantation regions. The confocal scanned area is marked on the selected electrode B2 near the top of two electrodes and the microwave strip line. The electrodes and microwave strip line pads are wire-bonded to the circuit board.
    (b) Optical and (d) photoelectrical confocal imaging of the region near the electrode exhibiting the strongest signal, resolving multiple isolated NV centers. Excitation was performed at 560\,nm (5\,mW, CW) with an applied bias of +3\,V.
    (c) Pulsed optically detected magnetic resonance showing characteristic $^{15}N$ isotope hyper-fine splitting $\sim$ 3.03\,MHz. 
    (e) Simplified photoelectric measurements scheme
    (f) Simplified scheme of laser induced (green arrows) spin-depended NV center conversion between neutral and negative charge states.}
    \label{fig:implanted_map}
\end{figure} 

\section{The setup and sample preparation}
Two samples were prepared: one with implanted NV centers and subsequent overgrowth following the process of Ref.\,\cite{findler2020indirect}, and one with natural ingrown NV centers.
Implantation was performed using $\mathrm{^{15}N^+}$ ions with an energy of 2.5\,keV and a dose of $\mathrm{1\times10^{11} N^+/cm^2}$ into the $\sim$50\,nm $^{12}$C layer grown on the E6 substrate (E6 [100] electronic grade). After implantation, the sample was overgrown with pure $^{12}$C layer for 30 min ($\sim$3--5 nm depth) and then annealed at 1000°C for 3 hours in vacuum \cite{balasubramanian2009ultralong, yamamoto2013extending}. The resulting depth of NV centers formed in the sample is expected to be $\sim$10\,nm. 
The expected impurity spin bath concentration is high, including both implanted nitrogen and crystal damage. The effect of overgrowth on the impurity content is discussed below and in Sec.~S.II of the Supplemental Material~\cite{SupplMat}. 

The second sample was used to examine the ingrown NV centers. The ingrown NV centers were found in the sample prepared by overgrowing (100 nm $^{12}$C) and annealing (1500°C) the E6 electronic grade diamond substrate. In both samples, the surface termination was oxygen. We fabricated electrodes $\sim 10\,\mu$m apart aligned with the implantation region. The electrodes were fabricated as described in \cite{bourgeois2015photoelectric}. Ti/Au metal layers were deposited in ultra-high vacuum with thicknesses of 30\,nm~(Ti) and 200\,nm~(Au). The electrodes have not been annealed. Perpendicular to them, a microwave strip line was placed in the same process and thickness. After lift-off, the pads of the structure were bonded to a printed circuit board (PCB) using 17.5\,$\mu$m gold wires. The same process is used to prepare electrodes on the sample with ingrown NV centers. 

The diamond with the metal structure for electrical detection is shown in Fig.\,\ref{fig:implanted_map}\,(a). As indicated in the figure, we perform measurements between the wire bonded electrode pair in the region close to the microwave line. The optical and electrical imaging was performed simultaneously using a home-built confocal microscope. The photoelectrical detection scheme is shown in Fig.\,\ref{fig:implanted_map}\,(e). The detection scheme consists of a bias voltage source connected to one of the electrodes, and the other electrode is used to collect the charge carriers. The signal from the collection electrode is amplified through the external current-to-voltage converter (also called a trans-impedance amplifier) with the gain as high as $\mathrm{10^{11}\,V/A}$ and the bandwidth of 200 Hz. The resulted voltage signal is digitized using a National Instruments Data Acquisition (NIDAQ) device. The resulting contact pair exhibits typical Schottky double diode behavior. To optimize charge collection and to increase the optimal SNR of the PE detection, the optimal bias voltage was chosen depending on the prehistory of applied bias. 

The scans shown in Fig.\,\ref{fig:implanted_map}\,(b)-optical and (d)-photoelectrical were made by scanning the focal point of the excitation laser and simultaneously measuring the photoelectrical and fluorescent signals (zigzag, left to right, bottom to top). The scan was performed with a laser excitation power of $\sim$5\,mW at 560\,nm (CW) in front of the air microscope objective (Olympus Apochromatic, NA\,=\,0.95), well above NV$^-$ optical saturation.
To suppress PE background contribution form $\mathrm{N_s}$, we tune the excitation wavelength to approximately 560~nm, corresponding to $\sim 2.21\,\mathrm{eV}$~\cite{hruby2022magnetic}. This wavelength remains sufficient to excite both NV$^-$ and NV$^0$ and to enhance the NV$^-$ population, while reducing direct ionization of $\mathrm{N_s}$~\cite{aslam2013photo}.
The bias was set to +\,3\,V. The bright spots on the confocal image correspond to the NV centers and show characteristic 2.87\,GHz resonance at zero bias magnetic field~\cite{gruber1997scanning, jelezko2002single}. The optical and electrical confocal images are in good agreement and the majority of spots found optically correspond to those detected electrically. The photoelectrical signal is the strongest close to the electrode and decaying with distance from it. Moreover, prolonged intense laser illumination can change the preferred charge state of the NV center more often, compare to ingrown NV centers. This appears as a reduced fluorescence rate, and as the reduced spin contrast discussed below. The background on the electrical scans is free of strongly correlated electrical grid noise such as 50 Hz and its harmonics. Some NV centers located very close to the electrode, has very strong photoelectrical signal, but are not very well distinguishable optically due to high parasitic fluorescence from the electrode. Most of NV centers found in this region show a hyperfine splitting of $\sim$3.03\,MHz which is characteristic of the $^{15}$N ions used for implantation~\cite{rabeau2006implantation}, confirming their artificial creation. The magnetic resonance spectrum is shown in Fig.\,\ref{fig:implanted_map}\,(c), acquired at low microwaves power (in order to avoid power broadening), using the conventional pulsed Optically Detected Magnetic Resonance (pODMR) sequence. The details of the pulsed measurements are described in the next section.

\begin{figure}[t]
    \includegraphics[width=\columnwidth]{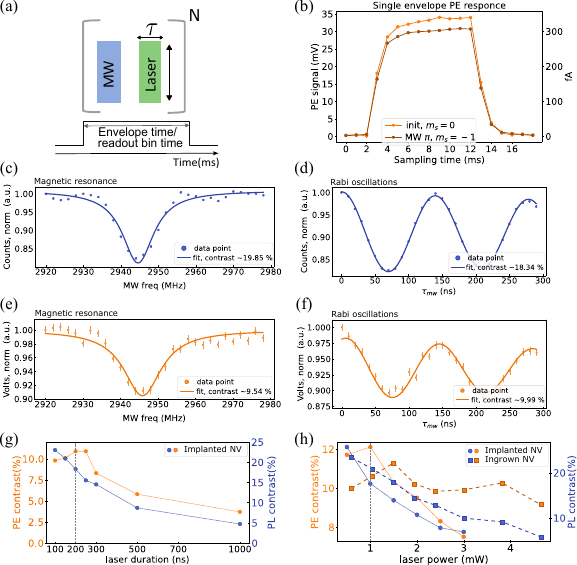}
    \caption{Pulsed photoelectrically detected magnetic resonance (PDMR) and Rabi oscillations. (a) Pulse envelope sequence utilized to detect spin-depended photocurrent. It consists of a microwave (MW) spin manipulation pulse followed by intense readout (spin initialization) laser pulse which is being optimized. (b) Measured averaged PE response from the pulse envelope. Initialization to $m_s=0$ without microwave pulses (bright curve) and by using microwave $\pi$ pulse to flip the spin into $m_s=-1$. (c) optically detected magnetic resonance and (d) corresponding Rabi oscillations. Photoelectrically detected (e) magnetic resonance and (f) corresponding Rabi oscillations. (g) PE(orange) and PL(blue) detected contrast at different readout laser pulse duration. (h) at optimum laser pulse duration (200 ns) contrast as a function of laser power. Circles and dashed line -- implanted and squares dotted -- ingrown NV centers.}
    \label{fig:pulsed_detection}
\end{figure} 

\section{Pulsed photoelectrical readout}
Next, we focused on characterizing the electronic spin properties of the NV centers. We used photoelectrical detection and compared the results with conventional optical detection. We performed the characterization starting with pulsed optically and photoelectrically detected magnetic resonance (ODMR/PDMR), followed by optically and electrically detected Rabi oscillations to determine the spin contrast.

The goal of any readout scheme (optical or photoelectrical) is to drive and detect emission whose magnitude depends on the spin state. In case of a NV center it is the electronic spin state formed by $m_s=0$ and $m_s=\pm1$ spin projections of $\mathrm{^3A_2}$ ground state (see Fig.\,\ref{fig:implanted_map}\,(f)). At applied static  magnetic field the degeneracy of $m_s=\pm1$ is lifted, allowing to drive selectively only one transition $m_s=(0,-1)$. In conventional optical readout, the NV center is driven below saturation of $^3$E excited state. Depending on the initial spin projection in the ground state the population exhibits inter-system crossing in the excited state $^3$E. It can either shelf in the long-lived ($\sim$ 300\,ns) singlet dark state for $m_s=-1$, producing fewer photons, or for $m_s=0$ relax back to the ground state producing more photons. Typically, a laser pulse of 3--5 $\mu$s for single NV centers is used for both spin readout and spin initialization into $m_s=0$. The resulting redshifted fluorescence of the leading edge is then detected with single photon avalanche photo diodes (APD) and digitized on a fast photon counting card. 
The PE readout exploits the same spin dependent singlet coupling. Following the widely accepted model in the literature \cite{siyushev2013optically}, the population is driven from the $\mathrm{^3A_2}$ ground state to the $^3$E excited state, where it has a high probability of decaying to the singlet if it is in the $m_s=-1$ state. If no singlet transition occurs, a second photon can ionize the system to the $^2$E ground state of the NV$^0$ charge state. This causes a free electron to be injected into the conduction band. To promote this process, the NV center is excited at elevated laser powers above saturation. From here, a third photon will transform the state into the $\mathrm{^2A_2}$ excited state of NV$^0$. Then a fourth photon will charge the system back to the ground state of NV$^-$, releasing a free hole. We assume that singlet states are shielded from photoionization, and thus the $m_s=0$ spin projection should correlate with a higher average carrier emission compared to $m_s=-1$. By measuring the resulted photocurrent, we can infer the electronic spin state from different control microwave pulses. If ionization from the singlet state is possible, this will lead to a reduced contrast of the detected spin-dependent photocurrent. 

We performed pulsed measurements in two ways. One method was performed at low laser power (below saturation) using a 5\,$\mu$s long laser pulse. The leading 300\,ns edge of the fluorescence is collected and digitized on the fast counting card. The second method uses high power laser pulses $\sim$ 200\,ns long for spin polarization and readout, with a waiting time between laser pulses of 1\,$\mathrm{\mu s}$. The microwave pulses for spin manipulation are applied between the laser pulses (see Fig.\,\ref{fig:pulsed_detection}\,(a)) and repeated $N \sim 10000$ times. The signal (photon count rate and electrical signal in Volts) is then acquired directly with the NIDAQ without any lock-in detection. 
The acquired PE signal from two spin states $m_s=0$ and $m_s=1$ is shown in Fig.\,\ref{fig:pulsed_detection}\,(b). It represents a single slow envelope response of N $\sim$ 8300 repetitions of laser and MW pulses with a characteristic amplifier rise/fall time of $\sim 2 ms$. The signal difference at the end of the envelope represents the PE signal difference of two spin states. We construct our readout sequence by measuring one bin per envelope at this point and varying the frequency of the MW pulse for pulse PDMR and the duration of the MW pulse for PE Rabi. The signal is then normalized to an envelope in which there are no MW pulses. 
The data is averaged by repeating the sequences up to thousand times. The first method is used to analyze only the fluorescence from the NV center and should produce almost no photocurrent (due to the low laser power). The second method produces spin-dependent photocurrent and fluorescence, which is detected simultaneously.

Fig.\,\ref{fig:pulsed_detection}\,(c) and (e) show optically and electrically detected magnetic resonance acquired with the pulse envelope method. The measurements were done at low static magnetic field aligned to the NV axis. On resonance, the corresponding Rabi oscillations were acquired with the same conditions, see Fig.\,\ref{fig:pulsed_detection}\,(d),(f). The presented PE detected magnetic resonance and Rabi oscillations are very similar to the optically detected ones. The main differences are the lower photoelectrical Rabi contrast and the asymmetrical wings of the PE signal. The latter is associated with the discharging of the electrodes in the beginning of the sweep sequence. This can be the reason why the PE detected magnetic resonance may appear wider than the PL detected. 
To mitigate this, several normalization pulses are appended to the beginning of the sequence, which removes parasitic charges from the electrode at the metal-diamond interface. These laser pulses are also initializing the spin state into $m_s=0$ state. In addition, we were sampling the PE signal slower than the amplifier's 2\,ms rise/fall time.
After that, the first acquired bins from the normalization pulses are cut out during data processing (see Fig.~S1 in the Supplemental Material~\cite{SupplMat})
With the normalization pulses, we swept the duration of the laser pulse (c) and laser power (d), presented in the Fig.\,\ref{fig:pulsed_detection}. The photocurrent (orange) contrast has an optimum laser duration of $\sim$ 200\,$ns$ and optimum laser power $\sim$ 1\,mW, measured as an average response from train of laser pulses with 1\,$\mathrm{\mu s}$ off time (vertical gray dashed line). 
As Fig.\,\ref{fig:pulsed_detection}\,(d) shows, there is a difference between shallow-implanted NVs, whose contrast quickly drops with increasing power, and ingrown NVs a few microns deep, whose contrast does not decay as strongly.
We attribute this to the Nitrogen bath which is significantly lower in the latter case. The magnetic resonance (c, e) and Rabi (d, f) plots in the Fig.\,\ref{fig:pulsed_detection} are displayed using the optimal parameters for laser power and duration for single implanted NVs.

\section{Electrically detected dephasing time}

\begin{figure*}[t]
    \includegraphics[width=0.85\textwidth]{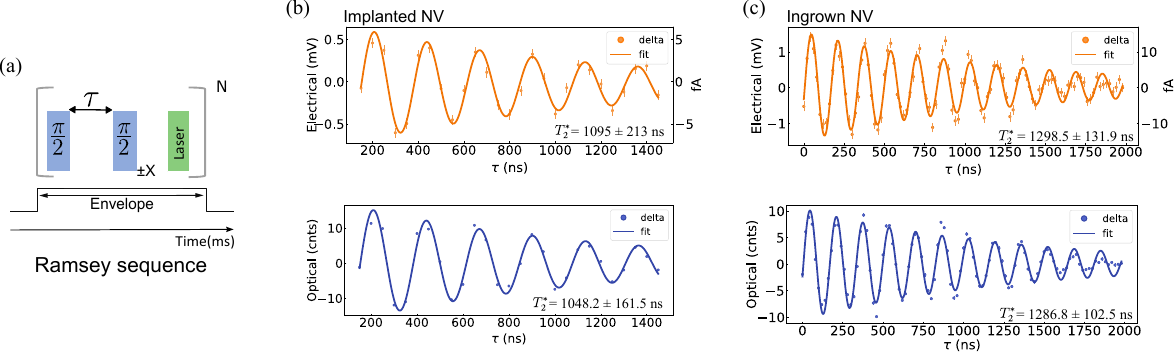}
    \caption{Electrical and optical Ramsey $T_2^*$ dephasing measurements of implanted and ingrown NV centers. Ramsey sequence encoded in the pulse envelope (a) where blue $\pi/2$ are microwave control pulses and green is short 200 ns polarization/readout intense laser pulse. The Ramsey sequence for implanted single (b) and ingrown NV centers (c).}
    \label{fig:PE_Ramsey}
\end{figure*} 

In this section we analyze dephasing of the electronic spin state of single NV centers detected optically and electrically. Here, we focused on the Ramsey interferometry protocol to measure $T_2^*$ dephasing times. Most of the measurements presented here are performed at a high magnetic field ($\sim$\,500\,G), around the excited state level anti-crossing (ESLAC), unless otherwise stated. We worked at this high field to ensure the polarization of the nitrogen ($^{15}$N or $^{14}$N) nuclear spin state, thus avoiding beating frequencies from hyper-fine sub-levels on the measured dephasing decay. In addition, the fluorescence and photocurrent signals become sensitive to the magnetic field misalignment with respect to the NV center axis, allowing precise alignment of the applied static magnetic field. 

The Ramsey protocol consists of two fast $\pi/2$ microwave pulses separated by phase accumulation time $\tau$ and consecutive laser readout pulse to measure the accumulated phase. The central frequency of the control pulses are detuned from the spin resonance frequency by $\sim$ 5--10\,MHz. We encoded such microwave control sequences in the pulse envelope to read out the spin state photoelectrically. For each $\tau$ we perform measurements with alternating of the phase ($+\text{Y}/-\text{Y}$) of the last $\pi/2$ control pulse. Resulted alternating Ramsey oscillations are subtracted from each other. An example of a Ramsey sequence block is shown in Fig.\,\ref{fig:PE_Ramsey}\,(a). The block is repeated N$\sim$10000 times, maintaining the fixed distance between laser pulses for different delays $\tau$ between $\pi/2$ pulses. The electrical and optical readouts are synchronized in the same way as for the pulsed PE-Rabi measurements discussed above. 
 
In the Fig.\,\ref{fig:PE_Ramsey}, Ramsey fringes are shown for implanted NVs (b) and ingrown NVs (c). The detection of PE (orange dots) and PL (blue dots) signals from pulse envelopes was performed simultaneously.  
The photoelectrically and optically detected Ramsey fringes, and the extracted $T_2^*$, agree for both implanted and ingrown NV centers, showing that the readout photocurrent does not measurably degrade the dephasing time.
The $T_2^*$ time does not change significantly at high or low laser power and is comparable within a standard deviation. In addition to the pulse envelope method, the $T_2^*$ decay for the ingrown centers is measured using conventional fast fluorescence readout with 5\,$\mu$s long low intensity laser pulses. This decay also reproduces very well the decay obtained by the pulse envelope method (see Fig.~S5 in the Supplemental Material~\cite{SupplMat}).

Typically, to increase SNR of the PE signal and to ensure efficient charge collection, the bias voltage is set in the order of several volts. However, we were able to detect the Ramsey fringes PE signal even at 0\,V bias. An example of such measurements on the same ingrown NV center is
shown in the Supplemental Material~\cite{SupplMat} (see Fig.~S5). Such a detection might be possible due to charge accumulation at the diamond-titanium interface. The interface can be easily charged due to the previously applied voltages. This can act as an effective bias field to the NV centers close to one of the electrodes and attract the electron/hole generated at the NV center. Understanding this effect may be important for improving charge collection and is beyond the scope of this manuscript.

\section{Discussions}
\subsection{The impact of the photocurrent on dephasing times}
A central question for photoelectric readout is whether the readout photocurrent itself limits the spin coherence. In both implanted and ingrown NV centers, we observe no significant influence of the photocurrent on $T_2^*$. 
This is consistent with Morishita et al.~\cite{Morishita2023}, who also observed no such influence in Hahn-echo $T_2$ measurements on NV ensembles using much higher photocurrents.
Moreover, the dephasing (e.g. DC Ramsey protocol) was measured in a pulsed regime where laser pulses (which produce the charge carriers) are separated in time. The charges produced by such laser pulses from the NV center are expected to be far enough away to significantly affect the dephasing time during the phase pickup between two adjacent $\pi/2$ pulses (for the Ramsey protocol). Any residual effect is expected to lie below the sensitivity of our measurement.
A parasitic field from moving surface charges could in principle influence the NV $T_2^*$, but the estimates of Acosta et al.~\cite{Acosta2022} give at worst a field of order $150$~pT. This is far below the reach of our apparatus, whose typical single-NV sensitivity is of order $\sim 1~\mathrm{\mu T/\sqrt{Hz}}$. We therefore do not expect the photocurrent to influence the dephasing times under our conditions.

While the readout photocurrent leaves $T_2^*$ intact, prolonged high-power laser illumination can shorten it through changes in the local charge environment.
Shallow NV centers are in any case sensitive to their surface: surface defects can reduce $T_2^*$~\cite{sangtawesin2019origins}, and the surface termination also affects the achievable dephasing time~\cite{fuhrmann2024probing}.
In one such case we observed shortening of the $T_2^*$ time after prolonged exposure of the NV center to high power laser pulses (see the Supplemental Material~\cite{SupplMat}). We observe such shortening from initial measurements at low laser power in implanted NV centers after long laser exposure (Fig.~S4).
Subsequently, after some hours the preferred negative charge state of the NV center was lost. We attribute this to the accumulation of charges in traps surrounding the NV center. 

A different $T_2^*$ shortening behavior was observed in both implanted and ingrown single NV centers at elevated laser power (above saturation) compared to the low laser power case. Measurements were performed using conventional fast fluorescence detection with 5 $\mu$s laser pulses and integration of the first $300$~ns fluorescence (see Figs.~S4--S6 in the Supplemental Material~\cite{SupplMat}).
The $T_2^*$ decreased faster in the implanted centers than in the ingrown centers. This may be related to the presence of a donor bath around the implanted centers, which is typically larger than for the ingrown NVs. The shortening might be also related to the laser-induced increased surface charge density from ionization of substitutional nitrogen or other defects.

Ramsey is the easiest protocol to read out photoelectrically, but it is only one of many for probing coherence. 
Dephasing time can be further improved by using pulsed schemes to decouple the spin state from the magnetic environment. These include more complex pulse sequences such as Hahn-Echo, \text{XY--8}, CPMG and others \cite{pham2012enhanced}. However, with the current state of photoelectric detection with pulse envelopes, it remains challenging. One of the main limiting factors is the photoelectric signal-to-noise ratio. As the microwave control pulse sequences become longer, the effective amount of charge generated by such an envelope becomes smaller. This is because the number of laser pulses decreases, resulting in low photoelectric response. By improving the charge collection and amplification, e.g. by gating each laser pulse, it could be possible to perform arbitrarily long control pulse sequences. This can be achieved, for example, by using fast high-resolution charge amplifiers maintaining high gain \cite{djekic2021440} or by creating an avalanche of each NV center photoionization event \cite{murooka2021photoelectrical}, similar to on-chip single-photon APDs.

\subsection{The impact of overgrowth on the PE background.}

Diamond overgrowth of the implanted layer is expected to modify the near-surface impurity content and the local defect environment~\cite{findler2020indirect, findler_PhDThesis_2026}. Vacancy-related complexes, substitutional nitrogen, and hydrogen-related defects can all contribute to the photoelectric background, i.e., to the photocurrent not originating from spin-dependent NV ionization. Nevertheless, a substantial fraction of implanted NV centers remains optically and photoelectrically active after overgrowth ~\cite{findler2020indirect}, enabling spin-dependent photocurrent detection.

A reduction of paramagnetic substitutional nitrogen after overgrowth is expected to suppress the background photocurrent associated with $\mathrm{N_s^0}$ ionization (see Sec.~S.II of the Supplemental Material~\cite{SupplMat}) and thereby improve the signal-to-noise ratio of the NV photocurrent. 
We measured the P1 concentration by Double electron--electron resonance (DEER) measurements~\cite{Grotz2011SensingExternalSpins} on a reference ensemble implanted with $2.5~\mathrm{keV}$ nitrogen ions at a dose of $1\times10^{13}~\mathrm{N^+/cm^2}$.
After overgrowth with an approximately 6 nm layer ~\cite{findler_PhDThesis_2026}, the P1 concentration dropped from $5.42\pm0.15~\mathrm{ppm}$ in the as-implanted, annealed sample to $2.07\pm0.11~\mathrm{ppm}$. 
The concentration of free electron spins with $g=2$ is also reduced, from $5.85\pm0.11~\mathrm{ppm}$ to $3.07\pm0.11~\mathrm{ppm}$~\cite{findler_PhDThesis_2026}. The decrease of the spin bath for lower implantation doses, used in this work is expected to be even more significant. Scaling the overgrown high-dose result to the implantation dose used in this work, $1\times10^{11}~\mathrm{N^+/cm^2}$, gives an estimated residual P1 concentration of approximately $20~\mathrm{ppb}$, which is probably below the sensitivity of DEER measurements on single NV centers~\cite{Li2021LocalDefectDensityDEER}.

Hydrogen incorporated during CVD overgrowth may additionally passivate electrically active defects or form hydrogen-related complexes~\cite{findler_PhDThesis_2026}, thereby modifying the local donor--acceptor balance in the implanted layer. Possible mechanisms include passivation of dangling bonds~\cite{Findler2024NVHNanoscale, Stacey2012HydrogenPassivationNV} at implantation-induced vacancy-related complexes or formation of NVH$^-$, H1$^-$, and H2-related centers~\cite{Goss2002HydrogenDiamond, Glover2004VacancyHydrogen, Mizuochi2004HydrogenRelatedDefects} (see Sec.~S.II of the Supplemental Material~\cite{SupplMat}). Such processes could reduce charge trapping and suppress background photocurrent. Therefore, reduced $\mathrm{N_s^0}$-related ionization together with hydrogen-assisted passivation provides a plausible interpretation of the improved photoelectric response. However, the microscopic origin of the background photocurrent reduction cannot be uniquely identified from the present measurements and requires further investigation, which is beyond the scope of this work.

\section{Conclusion}

In conclusion, we performed photoelectric readout of single shallow implanted NV centers in diamond and compared their response with that of a single deep in-grown NV center naturally occurring in the diamond substrate. In addition, we implemented photoelectrically detected coherent spin control using a Ramsey protocol and investigated the corresponding $T_2^*$ dephasing times. The photoelectrically detected $T_2^*$ agreed with optical readout and showed no measurable dependence on the readout photocurrent, indicating that the spin dephasing is preserved during electrical readout.
Photoelectric detection of implanted NV centers was possible despite the presence of a residual spin bath formed by substitutional nitrogen, hypothetical hydrogen-related defects, and implantation-induced defects after implantation and overgrowth. The overgrowth process is expected to improve the photoelectric background by passivating part of the spin bath that contributes to background photoionization, thereby enhancing the spin-dependent photocurrent response. 
These results represent an important step toward quantum sensing experiments and the realization of quantum registers based on deterministically generated single NV centers and coupled multi-NV systems. In such systems, photoelectric detection could enable selective electrical readout of individual NV spin states by placing electrodes near selected centers, potentially allowing readout below the diffraction limit of purely optical detection.

\section{Acknowledgment}
We thank the Ulm Center for Nanotechnology and Quantum Material for providing access to machines and the metal deposition. P.S. acknowledges funding from Baden-W\"urttemberg Stiftung and FWO via Odysseus project (Grant No. G0DBO23N). The work was supported by the German Federal Ministry of Research, Technology and Space (BMFTR) via future cluster QSENS and projects: DE-Brill (No. 13N16207), EXTRASENS (13N16935), DIAQNOS (No. 13N16463), quNV2.0 (No. 13N16707, DLR via project QUASIMODO (No. 50WM2170), Deutsche Forschungsgemeinschaft (DFG) via projects 386028944, 387073854, 445243414, 491245864, 499424854, 532771161, 546850640, and joint DFG/JST ASPIRE program via project 554644981, European Union's HORIZON Europe program via projects QuMicro (No. 101046911), SPINUS (No. 101135699), CQuENS (No. 101135359), QCIRCLE (No. 101059999) and FLORIN (No. 101086142), European Research Council (ERC) via Synergy grant HyperQ (No. 856432), IQST and Carl-Zeiss-Stiftung.

\bibliography{references_new}

\end{document}


\preprint{APS/123-QED}

\title{Photoelectrical readout and Ramsey interferometry of single shallowly implanted NV centers in diamond}

\author{Ilia Chuprina}
\email{ilia.chuprina@uni-ulm.de}
\affiliation{Institute for Quantum Optics, Ulm University, D-89081 Germany}

\author{Christoph Findler}
\affiliation{Institute for Quantum Optics, Ulm University, D-89081 Germany}
\affiliation{Diatope GmbH, Buchenweg 23, D-88444 Ummendorf, Germany}

\author{Johannes Lang}
\affiliation{Institute for Quantum Optics, Ulm University, D-89081 Germany}
\affiliation{Diatope GmbH, Buchenweg 23, D-88444 Ummendorf, Germany}

\author{Petr Siyushev}
\affiliation{Institute for Materials Research (IMO), Hasselt University, Wetenschapspark 1, B-3590 Diepenbeek, Belgium}
\affiliation{IMOMEC division, IMEC, Wetenschapspark 1, B-3590 Diepenbeek, Belgium}
\affiliation{Institute for Quantum Optics, Ulm University, D-89081 Germany}

\author{Fedor Jelezko}
\affiliation{Institute for Quantum Optics, Ulm University, D-89081 Germany}
\affiliation{Center for Integrated Quantum Science and Technology (IQST), 89081 Ulm, Germany}

\date{\today}

\maketitle

\onecolumngrid 

\section*{Supplementary Materials}
\renewcommand{\thefigure}{S\arabic{figure}}
\setcounter{figure}{0}  
\renewcommand\thesection{S:\Roman{section}}
\setcounter{section}{0}

\section{Experimental details}
\begin{figure*}[h]
    \includegraphics[width=0.4\textwidth]{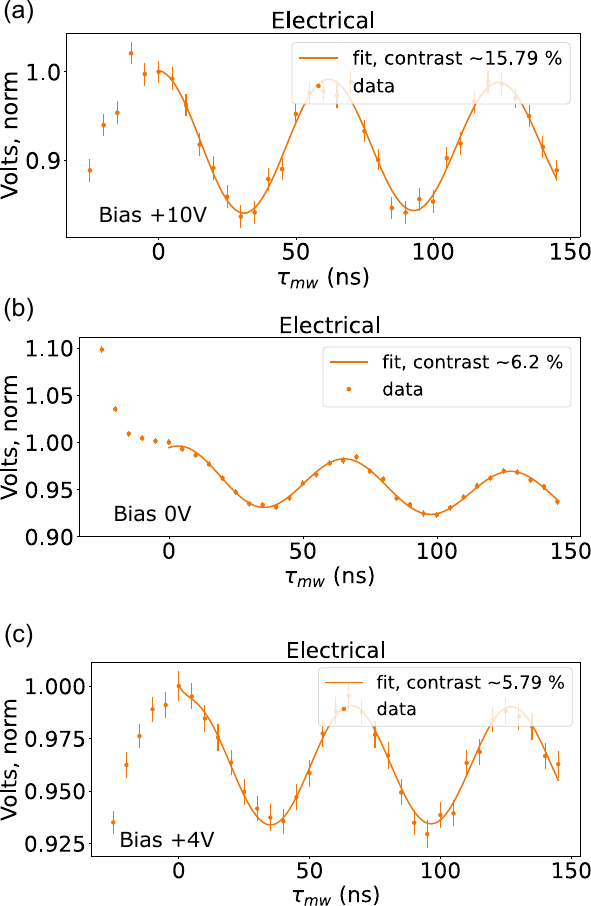}
    \caption{Photoelectrical Rabi on the same Ingrown NV under different Bias. Contrast is different depending on prehistory of bias voltage application.}
    \label{fig:figSM1}
\end{figure*}

Pulsed measurements (Rabi, pulsed PDMR, Ramsey) were performed using the pulse envelope approach due to the low detection bandwidth of the current amplifier (LCA-200-100G$\Omega$, rise/fall time 2 ms). The bias voltage is applied using a Keysight SMU in the typical range of -10 to +10 V. To perform Rabi type measurements, several envelopes were used without microwave (MW) control pulses (Tektronix AWG70001B) at the beginning of the Rabi pulse sequence. The electrical response of the first 5 envelopes at different bias voltages is shown in the Fig.\,\ref{fig:figSM1}. It is clearly seen that the electrical signal decay of the first envelopes is different at different voltages. The behavior is similar to the charge/discharge of a capacitor that can be formed in the diamond/metal interface. The interface can be charged by attracting charged carriers produced by ionization of the NV center and surrounding defects (such as substitutional nitrogen). Such space charge can significantly affect the electrical detection bandwidth and increase the rise/fall time of the detected electrical signals much more than 2 ms rise/fall time of the amplifier.
The laser power is measured on the photo-diode (Thorlabs S130C) as average response from envelope laser pulses of given duration and 1 $\mu s$ laser off time.

\begin{figure*}[h!]
    \includegraphics[width=1\textwidth]{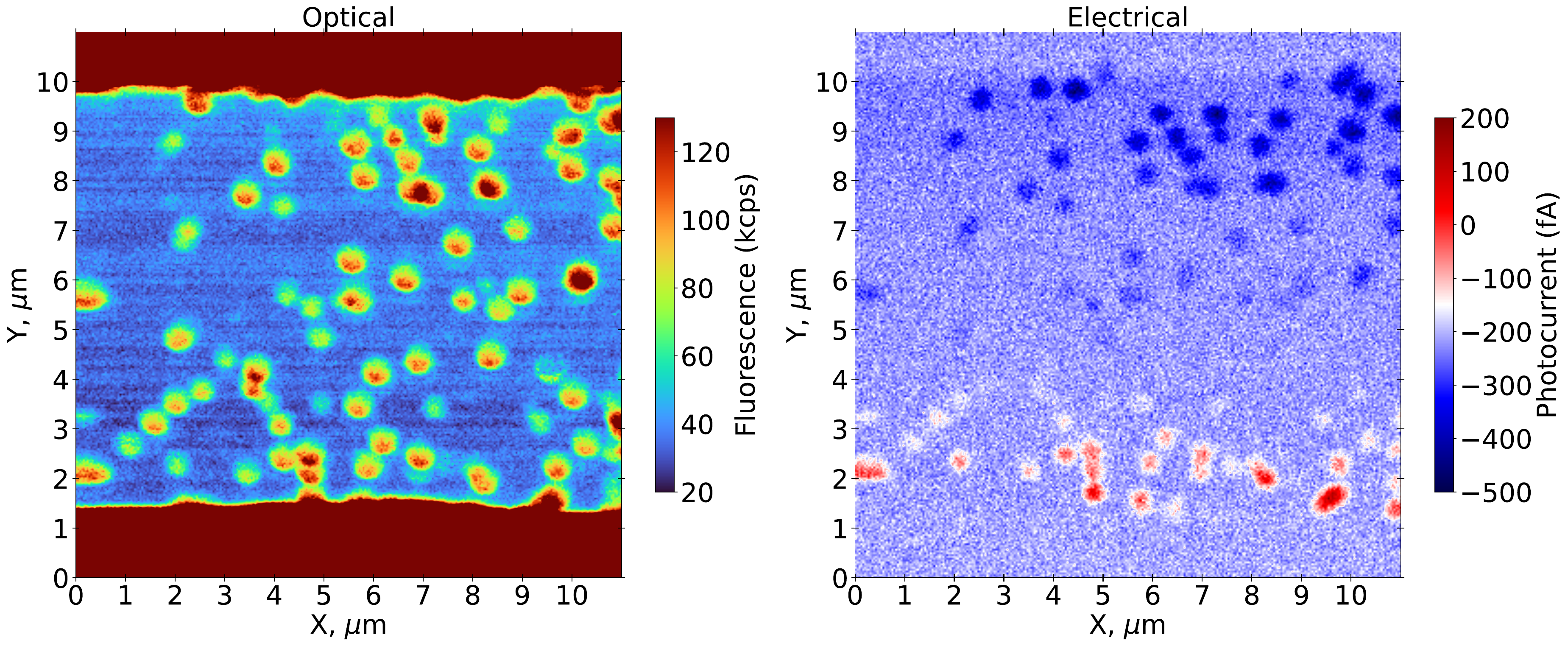}
    \caption{Imaging of the area between electrodes. Laser excitation 8\,mW at 560\,nm. Bias -3\,V, current collection from the bottom electrode. Two types of photoelectrical signal appeared: positive and negative. Imaging is done after long exposure and current signal collection from the top electrode.}
    \label{fig:figSM_imaging}
\end{figure*}

In addition, we regularly observed the charging of the interface after long measurements at elevated laser powers. High laser power promotes photoionization and trapping of the generated carriers by the electrodes, charging the interface and allowing photoelectric detection even at zero bias voltage. Interestingly, under negative voltage, we observed two types of signals from both electrodes, as shown in the Fig.\,\ref{fig:figSM_imaging}. The cycling of the bias voltage from positive to negative values and back leads to a charging of the interface, which causes hysteresis in the bias voltage and screening effects of the charged interface. Thus, it is possible to see only negative and only positive photoelectric signals produced by NV centers on the opposite side of the electrodes.

The bias history can significantly affect the optimal bias voltage due to hysteresis effects. However, even when the optimal bias has been selected, its value can change drastically over time, leading to a decrease in the observed SNR. We attribute this to the accumulation of charges on the electrodes during prolonged laser exposure.

\newpage
\section{Defect densities and photoelectric background in implanted and overgrown samples}

Ion implantation into diamond, including low-energy shallow implantation ($<5$~keV), inevitably produces lattice damage and introduces additional defect states~\cite{pezzagna2010creation,oforiokai2012spin,vanDam2019optical,favaro2017tailoring}.
Overgrowth with an ultrapure diamond layer can partially mitigate these effects by increasing the distance from the original surface and improving the spin dephasing of shallow implanted NV centers~\cite{staudacher2012enhancing,findler2020indirect}.
However, the overgrowth process may also modify the concentration of pre-existing defects or introduce new defect species. This is especially important for photoelectric readout of shallow, deterministically created NV centers, where carrier trapping, charge conversion, and photocurrent collection are sensitive to the local defect environment. Here, we discuss and estimate the concentrations of defect species relevant to improving photoelectric readout.
One of the main sources of photoelectric background is the ionization of substitutional nitrogen ($\mathrm{N_s}$) under visible laser illumination. Hydrogen-related defects, introduced during exposure to the hydrogen plasma in the CVD reactor, may also contribute to the local charge environment and photocurrent background. Since ion implantation and subsequent overgrowth can modify the defect landscape~\cite{findler2020indirect, stacey_APL2012}, the defect densities are expected to differ between implanted samples with and without overgrowth.
We estimate the average defect concentrations from theoretical depth profiles obtained for $2.5~\mathrm{keV}$ ion implantation using Crystal-TRIM simulations~\cite{Posselt01071994}. The resulting estimates are summarized in Table~\ref{tab:densities_crystal_trim}.

\begin{table}[h]
    \centering
    \caption{Defect densities in the three diamond configuration. (*From Crystal TRIM simulations \cite{Posselt01071994})}
    \label{tab:densities_crystal_trim}
    \begin{tabular}{p{4cm} c c c}
        \hline
        Configuration & Dose (ions/cm$^2$) & Nitrogen concentration (ppm) & Vacancy concentration (ppm) \\
        \hline
        Implanted nitrogen (2.5 keV) & 1e9 & 0.004* & 4.4* 
        \vspace{5pt}\\
        Overgrown implanted nitrogen (2.5 keV) & 1e11 & 0.373* & 435* 
        \vspace{5pt}\\
        In-grown nitrogen (annealed) & -- & $<\,0.005$ (E6 specs) & Na (probably very low) 
        \vspace{5pt}\\
        \hline
    \end{tabular}
\end{table}

Additionally, double electron--electron resonance (DEER) spectroscopy~\cite{Grotz2011SensingExternalSpins} with NV centers has been used to probe a local environment of nearby paramagnetic electron spins, including external spins, substitutional nitrogen defects, and other dark spin-bath constituents~\cite{DeLange2012MesoscopicSpinBath,Abeywardana2016NVESRSmallEnsemble}.
The spin-bath composition was estimated from DEER  measurements~~\cite{Li2021LocalDefectDensityDEER,Findler2024NVHNanoscale} on an NV ensemble implanted with $2.5~\mathrm{keV}$ nitrogen ions at a dose of $1\times10^{13}~\mathrm{N}^+/\mathrm{cm}^2$ (see C. Findler PhD Thesis~\cite{findler_PhDThesis_2026}). 

The overgrown sample shows a reduced concentration of the overall paramagnetic spin bath (Table~\ref{tab:bath_deer})~\cite{findler_PhDThesis_2026}. In particular, a reduced concentration of substitutional nitrogen, $\mathrm{N_s}$ (P1 centers), is expected to suppress the photoelectric background associated with P1 ionization.

For the implanted NV ensemble with $E=2.5~\mathrm{keV}$ and a dose of $1\times10^{13}~\mathrm{N}^+/\mathrm{cm}^2$, the P1 concentration extracted from DEER is $5.42~\mathrm{ppm}$ (Table~\ref{tab:bath_deer}), which is substantially lower than the value of approximately $40~\mathrm{ppm}$ expected from Crystal-TRIM simulations. This discrepancy indicates that only a fraction of the implanted nitrogen remains as paramagnetic substitutional nitrogen after implantation and post-processing, while the remaining nitrogen may be incorporated into other defect complexes, become electrically inactive, or be removed from the near-surface region during overgrowth and annealing. Since direct concentration measurements for the overgrown sample implanted at $1\times10^{11}~\mathrm{N}^+/\mathrm{cm}^2$ remain challenging, we estimate the expected P1 concentration by scaling the simulated implantation profile with dose. For the implantation dose used in the main-text experiments, $1\times10^{11}~\mathrm{N}^+/\mathrm{cm}^2$, this scaling gives an expected P1 concentration of approximately $20~\mathrm{ppb}$, based on the simulated value of $2.07~\mathrm{ppm}$ for a dose of $1\times10^{13}~\mathrm{N}^+/\mathrm{cm}^2$. Such a low concentration is difficult to resolve in DEER measurements on single NV centers.

\begin{table}[h]
    \centering
    \caption{Bath composition based on a DEER measurement~\cite{findler_PhDThesis_2026}}
    \label{tab:bath_deer}
    \begin{tabular}{p{4cm}cc}
        \hline
        Configuration 1e13 & P1 (ppm) & Free electron spins g=2 (ppm) 
        \vspace{5pt}\\
        \hline
        As-implanted, annealed & 5.42 +- 0.15 & 5.85 +- 0.11 \\
        \vspace{5pt}\\
        Overgrown with approx. 6 nm & 2.07 +- 0.11 & 3.07 +- 0.11 
        \vspace{5pt}\\
        \hline
    \end{tabular}
\end{table}

Although the amount of incorporated hydrogen cannot be determined from the present measurements, hydrogen-related defects are expected to be present due to exposure to the hydrogen plasma in the CVD reactor~\cite{findler_PhDThesis_2026}. These may include, for example, NVH, H1 ($\mathrm{VH}^0$), and H2-related centers~\cite{Goss2002HydrogenDiamond,Glover2004VacancyHydrogen,Mizuochi2004HydrogenRelatedDefects}. Such defects may modify the local charge environment and influence the concentration of electrically active donor- or acceptor-like states. Hydrogen may also interact with dangling bonds at implantation-induced vacancy-related defects, including through the formation of C--H bonds, which can modify the electronic structure and charge-state stability of vacancy-related complexes. Since these complexes can introduce electrically active states, hydrogen incorporation at such sites may affect charge trapping and the local donor--acceptor balance in the implanted layer. This can potentially passivate the spin bath around NV centers and improve their $\mathrm{T_2^*}$ dephasing properties (see Fig.~3 in Findler et.al. \cite{findler2020indirect}). However, the concentrations and depth distributions of hydrogen- and nitrogen-related defects are not necessarily comparable~\cite{findler_PhDThesis_2026} and require further clarification. Hydrogen incorporation can be substantial and strongly depth dependent, whereas the implanted nitrogen profile is localized near the surface \cite{findler2020indirect}. Consequently, hydrogen-related defects may be one possible contribution to the observed reduction of the photoelectric background, for example through passivation or modification of electrically active defects. This interpretation remains tentative, since the microscopic origin of the reduced background cannot be uniquely identified from the present data.

\section{Comparison to implanted NV centers without overgrowth}
\begin{figure*}[h]
    \includegraphics[width=0.8\textwidth]{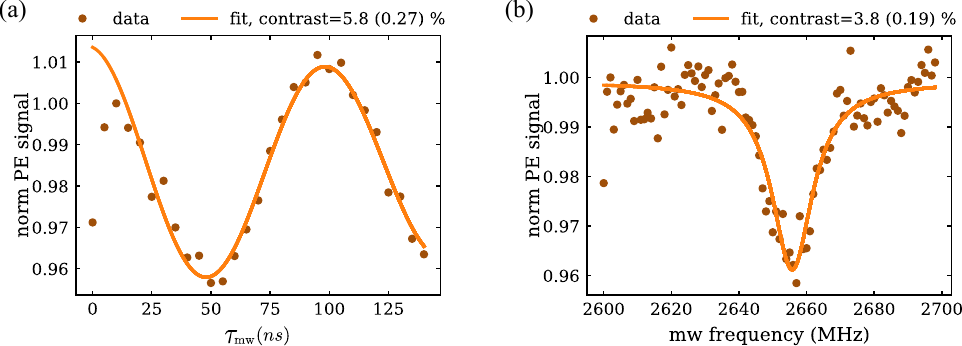}
    \caption{Typical photoelectrically detected Rabi oscillations on implanted NV center without overgrowth. Laser excitation wavelength 560 nm and power $\sim$5.5\,mW (CW). Bias +3\,V.}
    \label{fig:implanted_rabi}
\end{figure*}

We observed PE Rabi oscillations and pulsed PDMR signals from implanted NV centers without overgrowth. A representative signal is shown in Fig.\,\ref{fig:implanted_rabi}. In implanted and annealed samples, the background photocurrent is typically high and can be comparable to the photocurrent generated by an isolated NV center, resulting in a very low SNR in both CW and pulsed measurements. The PE background often increases further after bias-voltage cycling, which additionally degrades the SNR. As a result, it is frequently difficult to resolve the signal in pulsed measurement protocols that rely on slow envelopes.

At a fixed bias voltage, the PE background can be partially reduced by performing several slow confocal scans (4–10\,ms per pixel) focused at the depth of the NV center (close to the surface for implanted NVs) using high laser power, well above the NV saturation level, in the region near the collection electrode. We assume that this procedure depopulates nearby charge traps and helps stabilize the charge environment between the NV center and the collection electrode.

\section{Shortening of dephasing times}
\begin{figure*}[b]
    \includegraphics[width=0.9\textwidth]{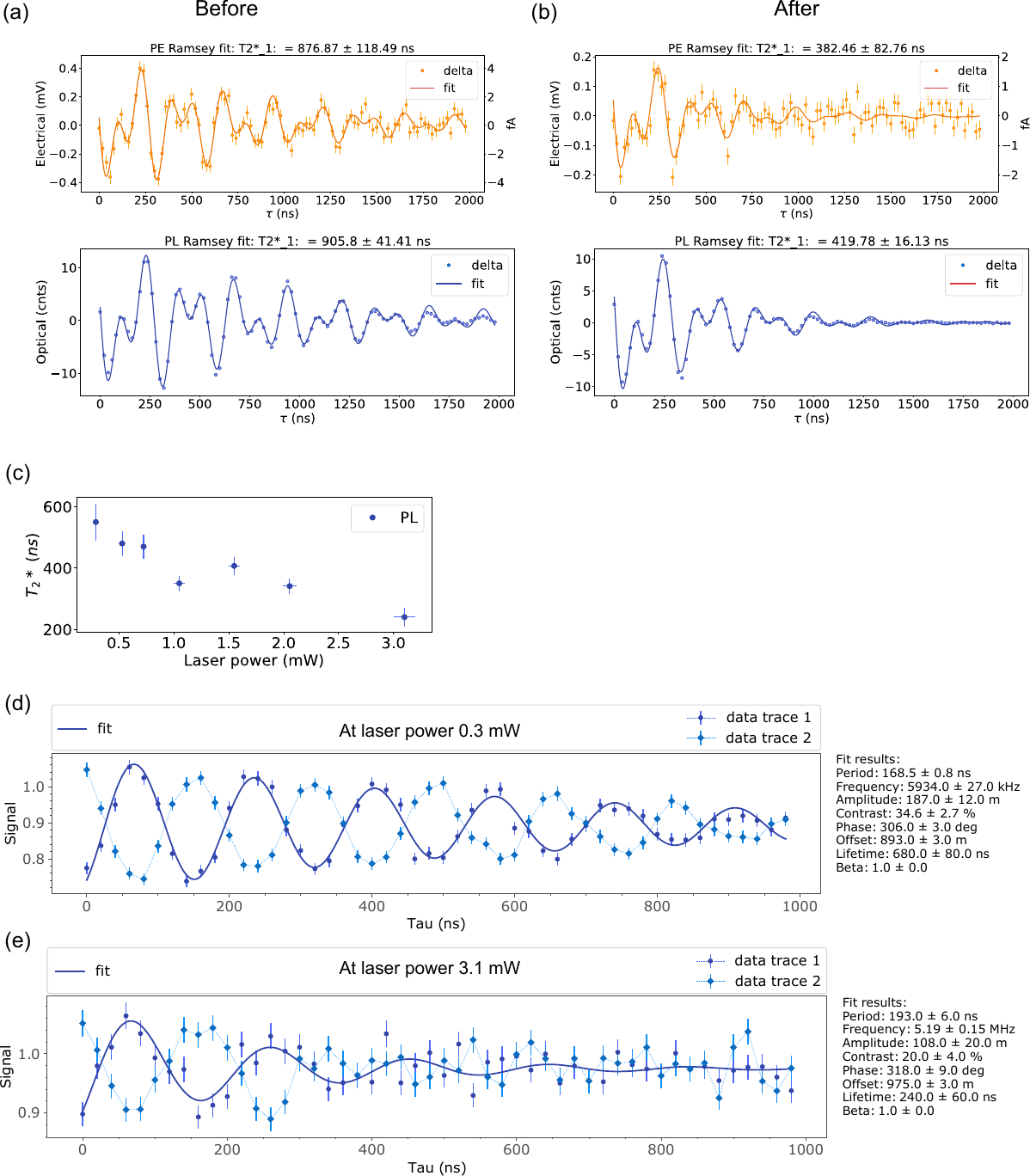}
    \caption{Shortening of Ramsey fringes lifetimes from single implanted NV center. Obtained using Pulse Envelope method conventional fast fluorescence detection. (a) Ramsey fringes at low magnetic field measured using pulse envelopes method and detuning 5\,MHz. Laser power 1 mW at 560 nm excitation wavelength, Bias +1\,V. (b) Ramsey fringes from the same NV center with shorter lifetime after long measurement with high power laser pulses.  Laser power 1 mW at 560 nm excitation wavelength, Bias 0\,V. (c) Ramsey fringes measurements via fast fluorescence detection at different laser powers. Shortening of lifetime for high intensity laser excitation. Corresponding Ramsey fringes at lowest 0.3 mW laser power (d) and highest laser power (e). }
    \label{fig:figSM3}
\end{figure*}

\begin{figure*}[h!]
    \includegraphics[width=0.9\textwidth]{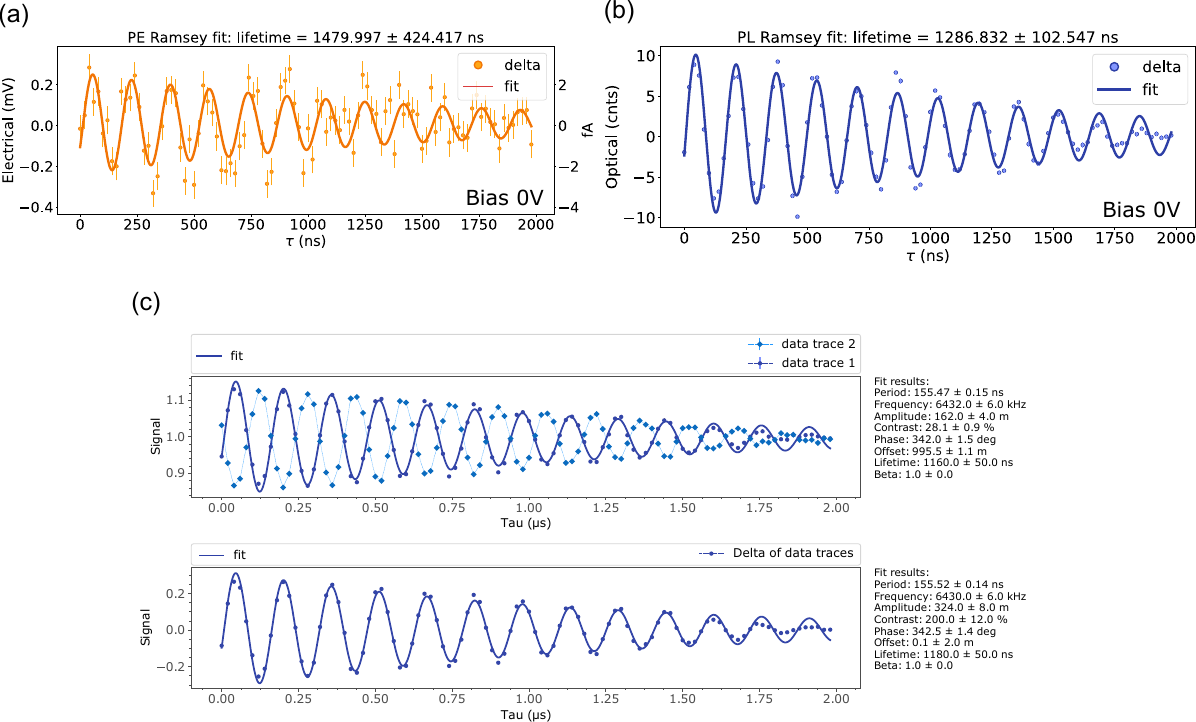}
    \caption{Additional Ramsey fringes. Ingrown NV same as in the main text. }
    \label{fig:figSM4}
\end{figure*}

\begin{figure*}[h!]
    \includegraphics[width=0.9\textwidth]{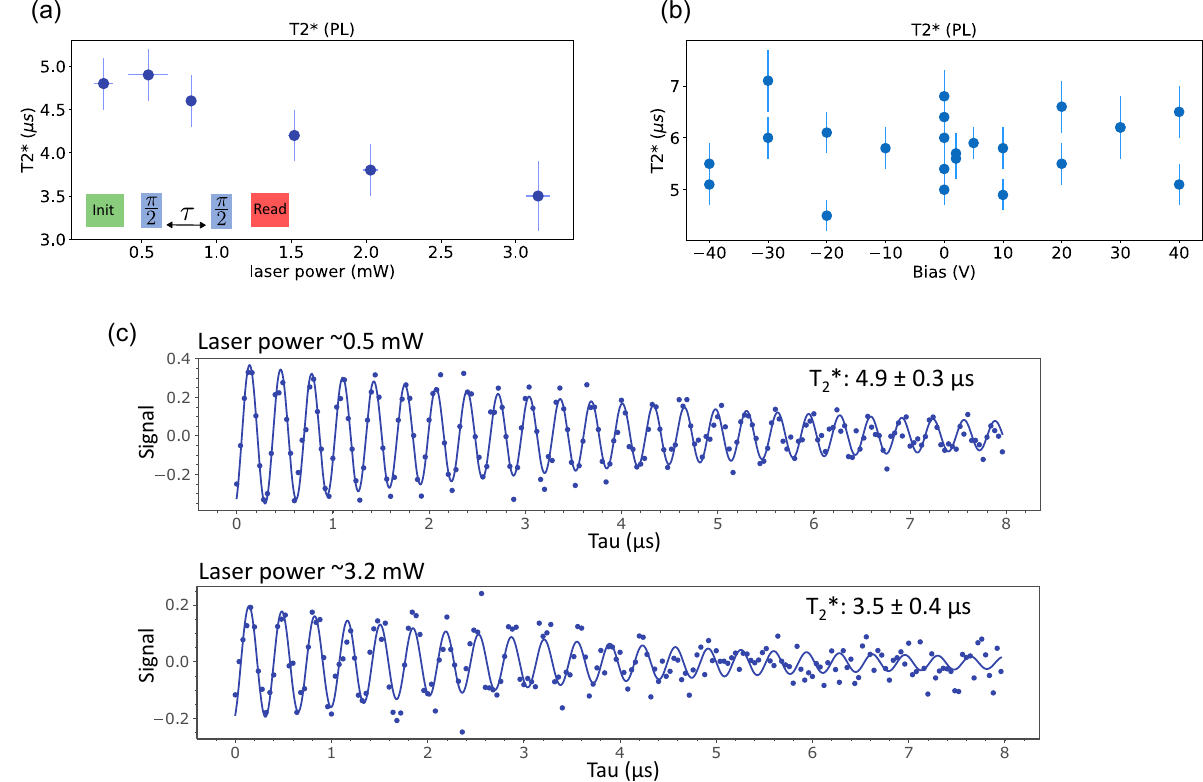}
    \caption{Additional data on $T_2^*$ Ramsey dephasing on ingrown NV centers. Shortening of $T_2^*$ at higher laser power.}
    \label{fig:figSM5}
\end{figure*}

A reduction of the dephasing time (Ramsey $T_2^*$) was observed for both implanted (with indirect overgrowth) and ingrown NV centers (see Figs.\,\ref{fig:figSM3}–\ref{fig:figSM5}). Measurements performed at increased laser power led to a shortening of the $T_2^*$ time. Repeating the measurements at lower laser power (typically below the NV saturation level) generally restored the dephasing time. However, for an implanted NV center we observed a drastic reduction of $T_2^*$ after prolonged laser illumination that did not recover afterward (see Fig.\,\ref{fig:figSM3}\,(a-b)).

\clearpage
\section*{References}
\bibliography{references_new}